\magnification=\magstep1
\baselineskip = 15pt
\def\hp{Hess and Philipp}
\font\title = cmbx10 scaled 1440 
\newcount\ftnumber
\def\ft#1{\global\advance\ftnumber by 1
          {\baselineskip=13pt \footnote{$^{\the\ftnumber}$}{#1 }}}
\newcount\fnnumber
\def\fn{\global\advance\fnnumber by 1
         $^{(\the\fnnumber)}$}

{\title
\centerline {Shedding (red and green) light on}
\medskip
\centerline{ ``time related hidden parameters''} 
}
\bigskip
\centerline{N. David Mermin}
\centerline{Laboratory of Atomic and Solid State Physics}
\centerline {Cornell University, Ithaca, NY 14853-2501}

{\narrower \narrower \baselineskip = 12 pt

\bigskip

\noindent I explain in elementary terms why the
critique of Hess and Philipp in Section 3.2 of quant-ph/0103028 fails
to invalidate the nontechnical version of Bell's theorem I gave twenty
years ago, involving two detectors with 3-pole switches and red and
green lights.

}

\bigskip

In 1985\ft{N. David Mermin, Physics Today, April 1985, 38-47.} I
described a very simple special case of Bell's theorem in which two
far-apart detectors each have three possible settings (labeled 1, 2,
and 3), randomly and independently selected after two appropriately
correlated particles have left their common source, and before either
particle has arrived at its detector.  When a particle does arrive its
detector flashes a red (R) or green (G) light. The data accumulated in
many runs of this experiment have two important features:\ft{Although
it is irrelevant to any of the points made below except for footnotes
4 and 13, I note that such data are produced by two spin-1/2 particles
in the singlet state, when the detectors are Stern-Gerlach magnets,
the three settings are associated with measuring the spin along three
coplanar directions 120$^\circ$ apart, and R and G signal spin-up and
spin-down at one detector, while signalling  spin-down and spin-up at
the other.}

(i) In those runs in which the detectors happen to have been given the
same settings, the light always flash the same color.

(ii) If all runs are examined without reference to the settings of 
the detectors, the pattern of flashes is completely random; in
particular the colors flashed are equally likely to be the same or
different.  

To account for feature (i) in the absence of any communication between
the two wings of the experiment it is plausible to entertain the
hypothesis that in each run both particles carry to their detectors
identical sets of instructions specifying what color the detector is
to flash for each of the three possible settings.  The instruction set
GGR, for example, means flash G for settings 1 and 2, R for setting 3.
An instruction is required for every one of the three possible settings,
because the settings are not chosen until after the particles have
separated, and each of the pairs of settings 11, 22, and 33 has 1/9 of a chance
of being chosen in any given run.  The particles must carry such
identical instruction sets in every single run of the experiment, since each
run has a 1/3 chance of ending up one in which the settings are the
same.

But this obvious explanation of feature (i) is incompatible with
feature (ii) of the data.  Since each of the nine possible pairs of
settings is equally likely in any run, any of the six instruction sets
in which both colors appear (for example GGR) will result in the same
color flashing 5/9 of the time (in 11, 22, 33, 12, and 21 runs).  The
remaining two instruction sets, RRR and GGG, result in the same colors
flashing every time.  Consequently if such instruction sets were the
correct explanation for feature (i), then when the data was examined
without regard to the settings, the same colors would be found to
flash at least 5/9 of the time, contradicting feature (ii).\ft{Bell's
inequality in this simple example is ``same at least 5/9'' of the
time''.  It is violated by the data: ``equally likely to be the same
or different''.  Note that this argument is independent of the
specific form of the distribution of instruction sets, which can vary
from one run to the next (and can even depend on all the data collected in
all earlier runs).  This is not, however, the kind of time dependence
that \hp\ want to pin their local hidden-variables model on.}  One is
left with the puzzle of how to explain feature (i), in the absence of
such instruction sets.\ft{Quantum mechanics does, of course, provide an
explanation, if you want to call it that --- namely the one given in footnote
1 above.}

I first put forth this special case of Bell's theorem over twenty years
ago\ft{N. David Mermin, American Journal of Physics {\bf 49}, 940-943
(1981).} to demonstrate to nonscientists in a simple but rigorous way
precisely what was so extraordinary about quantum correlations.  In the
intervening years I have found that its transparency also makes it a
good testing ground for claims of conceptual error in the formulation
or proof of Bell's theorem.  Confusion buried deep in the formalism of
very general critiques tends to rise to the surface and reveal
itself when such critiques are reduced to the language of my very
elementary example.

Such claims of error have recently been made on arXiv\ft{Karl
Hess and Walter Philipp, quant-ph/0103028, quant-ph/0206046.} and
elsewhere\ft{Proc. Nat. Academy Sci. (USA) {\bf 98}, 14224-27,
14227-34 (2001); Europhys.~Lett.~{\bf 57}, 775-781 (2002).} by Karl
Hess and Walter Philipp.  Conveniently, in Section 3.2 of
quant-ph/0103028 they bring their elaborate general argument to bear
on my 1985 version of Bell's theorem.  In that context their criticism
of Bell's theorem is so simple that it is easy to explain why it fails
to undermine my argument that no local classical instruction
sets can account for the data.

According to \hp\ my argument overlooks the fact that associated with
each of the three randomly and independently chosen settings of each
detector in each run of the experiment can be an enormous number of
different possible microscopic settings, one of which, varying from
one run to the next, describes the actual condition of the detector in
any given run.  Even though the {\it settings\/} are randomly and
independently chosen in each run, the underlying {\it microsettings\/}
for a given pair of settings can be correlated between the two
detectors because, for example, of time-dependent but common
conditions prevailing at the two detectors at the moment\ft{All
temporal statements may be interpreted in a given inertial frame of
reference --- for example the laboratory frame.}  they are
triggered.\ft{From this point onwards the reader must take care always
to distinguish between the settings (1, 2, or 3) and the microsettings
that underly them.} My simple three-entry instruction sets can
therefore be expanded to much larger sets that tell each detector how
to flash for each of the many possible microsettings that may be
underlying each setting.  Because the microsettings that underlie two
settings can be correlated even though the settings themselves are
independently random,
\hp\ assert that my argument ``cannot proceed.''

They do not, however, spell out {\it why\/} it cannot proceed in the
face of this complication.  Nor do they produce a local classical
hidden-variables model, tailored to my simple special case, that
exploits correlations of the microsettings to produce the data I
describe.\ft{They do present a very general hidden-variables model
intended to reproduce the quantum correlations in the singlet state
for the whole continuum of possible settings.  Myrvold
(quant-ph/0205032) has adapted their very elaborate construction to
the special case of only two settings at each detector where he shows
that it is explicitly nonlocal.  The same conclusion is reached more
generally by R.~D.~Gill, G.~Weihs, A.~Zeilinger, and M.~Zukowski (private
communication).}  The reason for the latter shortcoming is that my
argument can, in fact, proceed, with very little complication.
Expanded instruction sets cannot do the job.

What \hp\ fail to note is that if instruction sets are to
be based on the microsettings that underly the settings, then the
requirement (i), that the lights flash the same colors when the
settings are the same, enormously constrains both the possible
correlations between the microsettings at the two detectors, and the
possible forms of the expanded instruction sets.

To see this note first that if every pair of microsettings associated
with the same two settings (11, 22, or 33) has a non-zero probability
of occurring in a run, then no matter how strongly correlated the
microsettings may otherwise be, the expanded instruction sets for that
run must assign the same color to every microsetting, to guarantee
that the lights flash the same colors when the settings are the same.
Evidently it is enough to specify what that common color is for each
of the three entire collections of microsettings --- i.e. for each of
the three settings.  The expanded instruction sets of
\hp\ then collapse back to the instruction sets of my example.  As before,
these must exist in every run, whether or not the detectors do end up
with the same setting, because every run has probability 1/3 of being
either a 11, 22, or 33 run, and the particles have to be prepared for
each of these possibilities when they leave their source.  The
impossibility of instruction sets then follows exactly as it does when
microsettings are not explicitly taken into account.

For more general pair distributions of microsettings, there can be
instruction sets which assign more than just one color to all the
microsettings that can accompany a given setting, but because of
feature (i), this is only possible if the microsettings are strongly
correlated in a very particular way.  Microsettings associated with
the same setting must be constrained so that a microsetting at one
detector, instructed to flash red, cannot coexist with a microsetting
at the other, instructed to flash green.  This requires the
microsettings for each setting and at each detector to fall into one
of two distinct types, I and II, such that for each setting the
underlying microsettings at the two detectors can only coexist with
those of the same type.  Since the settings need not be chosen until
just before the particles arrive at their detectors and since any run
has a 1/3 chance of ending up with the same settings, these divisions
of the microsettings into two types must hold in every run and for all
three settings.

So as
\hp\ emphasize, the instruction sets 
can indeed be more elaborate than the ones I describe in Ref.~1.  But
their extension is highly constrained.  The most general possible
extended instruction set is restricted to specifying for each setting
whether type I microsettings flash red and type II green, or vice
versa.\ft{Should all microsettings for a given setting be of the same
type, call it type I.}
This again takes us back to the instruction sets of my paper, though
now we have to reinterpret them.

Now GGR in a given run means that the type-I microsettings under
settings 1 and 2 result in a green flash and the type-I microsettings
under setting 3 result in a red flash, while the type-II
microsettings for each setting flash the color opposite to that
specified by the instruction set. If particles with instruction set
GGR arrive at their detectors and discover, for example, that the
(common) conditions prevailing at the detectors at that moment are the
kind that require type-II microsettings for settings 1 and 3 and
type-I microsettings for setting 2, then the ouput produced by that
instruction set for each of the nine possible pairs of settings will
be that specified by the instruction set RGG of my original
uncomplicated model.  Since each of the nine pairs of settings are
equally probable, this effective instruction set will once again
result in the same colors flashing 5/9 of the time.  Whatever the
reinterpreted instruction set turns out to be, it will either result
in the same colors flashing 5/9 of the time, or (if it is GGG or RRR)
all the time.  We are back again to the argument of my paper.

The only complication introduced by the microsettings of \hp\ is that
the data actually produced by each 3-entry instruction set is only
determined when the particles arrive in the neighborhood of their
detectors and learn the particular character of the conditions
prevailing at the detectors.  This has no effect on the validity of
the argument that (extended) instruction sets capable of producing
feature (i) of the data cannot be compatible with feature
(ii).\ft{Even this minor complication is not really necessary. Since
the conditions prevailing at the detectors (for example the time on
two synchronized clocks) must have identical predetermined
consequences for whether type-I or type II microsettings underly each
of the three possible settings at the two far-apart detectors, it
would be a very strange model of the world that did not make the same
information (synchronized clock time) directly available at the source
as well thereby permitting the instruction sets to assume their final
forms before the particles left the source.  (The source must know how
long each particle takes to reach its detector, since they must arrive
at the same time.)}$^,$\ft{It is also worth remarking that according
to quantum mechanics the statistical character of the data in an EPR
experiment is unaffected if the two detections are separated by
arbitrarily long time intervals, provided no interactions intervene to
disrupt the singlet-state spin correlations.  To maintain this feature
of quantum mechanics in a \hp\ embellishment of my model it would be
necessary for the choice of type I or type II microsettings for each
setting to be the same for all times. The information about that
choice could be available at the source, taking us back to precisely
the instruction sets of Ref.~1, without even the modest embellishment
described above.}

I conclude that the reason \hp\ have not completed the critique in
Section 3.2 of quant-ph/0102038 with a counterexample tailored to that
special case is that none exists.  My argument against an explanation
based on instruction sets is easily expanded to accomodate their
``time related hidden parameters''.\ft{Since their very general and
elaborate local hidden-variables model based on time-related hidden
detector parameters includes my elementary example as a special case,
it too must contain errors, as has been argued directly in
Refs.~10.}

\bigskip

{\sl Acknowledgments.} I thank Richard Gill, Gregor Weihs, and Marek
Zukowski for helpful comments.  Supported by the National Science
Foundation, Grant No.~PHY0098429.

\bye